\documentclass[prb,preprint,showpacs,showkeys]{revtex4}
\usepackage{graphicx,bm,texdraw,epsf,amssymb,amsbsy}
\newcommand{\ve}{\varepsilon}
\newcommand{\be}{\begin{equation}}
\newcommand{\ee}{\end{equation}}
\newcommand{\bea}{\begin{eqnarray}}
\newcommand{\eea}{\end{eqnarray}}
\begin{document}
\title{A massive Feynman integral and some reduction relations for Appell functions}
\author{M.\ A.\ Shpot}\thanks{Electronic mail: shpot@ph.icmp.lviv.ua}
\affiliation{Institute for Condensed Matter Physics, 79011 Lviv, Ukraine}
\affiliation{Fachbereich Physik, Universit{\"a}t Duisburg-Essen,
D-47048 Duisburg, Germany}
\begin{abstract}
New explicit expressions are derived for the
one-loop two-point Feynman integral with arbitrary external momentum and
masses $m_1^2$ and $m_2^2$ in $D$ dimensions.
The results are given in terms of Appell functions, \emph{manifestly symmetric}
with respect to the masses $m_i^2$.
Equating our expressions with previously known results
in terms of Gauss hypergeometric functions
yields reduction relations for the involved Appell functions
that are apparently new mathematical results.
\end{abstract}
\pacs{11.10.Kk, 12.38.Bx, 02.30.Gp, 05.70.Jk}
\keywords{Feynman integrals, Appell functions, hypergeometric functions,
field theory, critical behavior, Lifshitz point}
\maketitle
\section{Introduction}\label{Intro}
In this paper, we derive simple new analytic expressions for a
standard $D$ dimensional one-loop two-point Feynman integral with
two different masses $m_1^2>0$ and $m_2^2>0$ and an arbitrary external momentum
$\bm p_x$,
\be\label{Idef}
I(p_x;m_1^2,m_2^2)=\int\frac{d^Dp}{(2\pi)^D}\;
\frac{1}{p^2+m_1^2}\;\frac{1}{(\bm p+\bm p_x)^2+m_2^2}\;.
\ee
Our results are given terms of a single, manifestly symmetric
Appell function \cite{Appell26} $F_4$, or alternatively,
the Appell function $F_1$.

The integrals like that in (\ref{Idef}) are ubiquitous in quantum
field theory and renormalization group theory of phase transitions
and critical phenomena. Actually, there exists for long time a
standard general result for the integral (\ref{Idef}) given by Berends,
Davydychev, and Smirnov in Ref. \onlinecite{BDS96} henceforth
called BDS. This result was expressed essentially in a form of a
linear combination of two Gauss hypergeometric functions of
complicated arguments.

Clearly, the Feynman integral (\ref{Idef}) is a symmetric function
of its parameters $m_1^2$ and $m_2^2$. In the result of BDS, which
we shall quote further, this property was implemented through a
symmetrization
\be\label{Sympr}
F(m_1^2,m_2^2)+F(m_2^2,m_1^2)
\ee
of two complicated terms, none of which obeyed the symmetry
$m_1^2\leftrightarrow m_2^2$.

Here we present the alternative general results for the integral
(\ref{Idef}). These are given in terms of a single Appell
function, $F_4$ or $F_1$, \emph{manifestly symmetric} with respect
to the interchange of $m_1^2$ and $m_2^2$. The symmetry
$m_1^2\leftrightarrow m_2^2$ is
implemented within the arguments of these functions. That is, our
new expressions are given in terms of completely symmetric
combinations like $m_1^2m_2^2$ or $(m_1\pm m_2)^2$. They do not
contain any symmetrizations like that in (\ref{Sympr}).

We believe that such kind of results are interesting on their own
right, and hope that their existence can be useful in the quantum
field theory. Moreover, comparing our results with the expression
of BDS we arrive at special identities between the Appell and
Gauss hypergeometric functions that are apparently new
mathematical relations.

Moreover, there exists another reason for practical utility of our
results. In fact, our motivation for present calculations was in
the theory of the critical behavior of statistical mechanical
systems at Lifshitz points \cite{HLS75}. Reviews and extensive
lists of references on this topic can be found in Refs.
\onlinecite{Hor80,Sel92,Die02,Die05}.

The field theoretical treatment of $m$-axial Lifshitz points in
$d$-dimensional systems frequently requires calculations of
momentum integrals
\bea\label{LPint}
&&J(p_x,q_x)=
\\&&\nonumber
\int\frac{d^mq}{(2\pi)^m}\int\frac{d^{d-m}p}{(2\pi)^{d-m}}
\frac{1}{ (\bm p-\bm p_x /2)^2+(\bm q-\bm q_x /2)^4}\,
\frac{1}{ (\bm p+\bm q_x /2)^2+(\bm q+\bm q_x /2)^4}
\eea
in contexts of the $\ve$-expansion
(see e.g. Ref. \onlinecite{SD01}), or large-$N$ expansion \cite{SPD05}. In this
setting, an $m$-dimensional subspace corresponding to $0\le m\le
d$ special anisotropy directions must be split out within the
whole $d$-dimensional space and complemented by the remaining
$(d-m)$-dimensional subspace with no peculiarities in its physical
properties. The momentum integral in (\ref{LPint}) is over
${\mathbb R}^d=\mathbb{R}^m\oplus\mathbb{R}^{d-m}$, and the inner
integration over $\bm p$ in $\mathbb{R}^{d-m}$ is easily recognized to be the
same as in (\ref{Idef}) with $D\equiv d-m$ and identifications of
masses $m_1^2$ and $m_2^2$ as $(\bm q\pm\bm q_x/2)^4$. In the
fully isotropic limit $m=0$, we recover the usual situation of the
Euclidean field theory in $D=4-\ve$ dimensions with the standard
upper critical dimension $D^*=4$. The latter follows from
$d^*(m)=4+m/2$, the upper critical dimension appropriate for the
Lifshitz-point theory, as the number of the anisotropy axes $m$
vanishes.

In the generic case of non-isotropic $m$-axial Lifshitz points
with $0<m<d$, when we are interested, for example, in the one-loop
energy-energy correlation function, a further integral over $\bm
q$ has to be still performed in (\ref{LPint}). To be feasible, its
integrand must be as simple and symmetric as possible. We could
not use the BDS result of the inner $\bm p$ integration to proceed. The
reason was the lack of symmetry $m_1^2\leftrightarrow m_2^2$ in
individual terms $F$ of (\ref{Sympr}). This implied the presence of
complicated explicit dependencies on \emph{odd} powers of the
integration variable $\bm q$ in each of these terms. Of course,
they cancel out in the symmetric sum (\ref{Sympr}), which is an
{\em even} function of $\bm q$ (and $\bm q_x$). But the outer
integration of any part of this sum appeared completely hopeless.

In a hope to achieve the desired simple symmetry of the inner
integral in (\ref{LPint}), we have differently tried to apply
the numerous standard transformation formulas
of Gauss hypergeometric functions in the BDS result, but in
vain. The application of the Feynman-parameter integration
also lead us only to results equivalent to that of BDS. Hence, we
were urged to attempt a completely different calculation in order
to get the inner integral given separately by (\ref{Idef}) in a simplest
possible and manifestly symmetric form.

Before turning to details of our calculation and its implications,
we write down the BDS result in the following section.

\section{The result of Berends, Davydychev, and Smirnov  \cite{BDS96}}

The Feynman integral of equation (\ref{Idef}) obeys the scaling
relation
\be\label{Scaling}
I(p_x;m_1^2,m_2^2)=p_x^{-\ve}\;I(1;m_1^2/p_x^2,m_2^2/p_x^2)
\ee
where we define, as usual for the critical phenomena theory, $D=4-\ve$
(note that this differs from the space-time dimension $n=4-2\ve$ of BDS).
Hence, without loss of
generality we can consider the function (cf. (\ref{LPint}))
\be\label{I1q}
I(m_1^2,m_2^2)\equiv
I(1;m_1^2,m_2^2)=\int\frac{d^Dp}{(2\pi)^D} \frac{1}{ (\bm p-\bm 1/2)^2+m_1^2}\;
\frac{1}{(\bm p+\bm 1 /2)^2+m_2^2}
\ee
where $\bm 1$ denotes an arbitrary vector of unit length. The dependence of
the original integral (\ref{Idef}) on external momentum $p_x$
can be easily reconstructed by scaling at any stage of
calculations.

The reference \onlinecite{BDS96} contains the explicit result for this
integral in terms of Gauss hypergeometric functions. Before
quoting it, let us split from the function $I(m_1^2,m_2^2)$ an
overall numeric factor via
\be\label{Geomf}
I(m_1^2,m_2^2)=(4\pi)^{-\frac{D}{2}}\Gamma\big(2-\textstyle{\frac{D}{2}}\big)
\,\hat I(m_1^2,m_2^2)\,.
\ee
While $\Gamma(2-D/2)$ is singular as
$D\to 4$, the non-trivial function $\hat I(m_1^2,m_2^2)$ is finite
in this limit. In  our conventions, the translation of the BDS
result (see Eq. (A.7) of Ref. \onlinecite{BDS96}) reads
\bea\label{Ihat}
&&\hat I(m_1^2,m_2^2)=\frac{  \Gamma^2(D/2-1)  }{  \Gamma(D-2)  }
\sqrt\Delta ^{D-3}
\nonumber\\&&\nonumber
+\frac{m_1^{D-4}}{D-2}\,(1+m_2^2-m_1^2-\sqrt\Delta)\, _2F_1\Big(
1,2-\frac{D}{2};\frac{D}{2}; -\,\frac{
(1+m_2^2-m_1^2-\sqrt\Delta)^2 }{ 4m_1^2 } \Big)
\\&&
+(m_1^2\leftrightarrow m_2^2)\,.
\eea
For brevity, we denoted by
$(m_1^2\leftrightarrow m_2^2)$ the presence of the third term,
repeating the function given explicitly in the second line, but
with interchange of the masses $m^2_1$ and $m^2_2$. The symmetric
combination
\be\label{Mathcal}
\Delta=(m_1^2+m_2^2+1)^2-4m_1^2m_2^2=\big[(m_1+m_2)^2+1\big]\big[(m_2-m_1)^2+1\big]
\ee
is a counterpart of the K\"allen function with unit momentum
in the terminology of Ref. \onlinecite{BDS96}. On the other hand, this is the
discriminant of a quadratic equation that has to be solved in the
course of calculations employing the Feynman parametrization.
The apparent singularities at $D=2$ in (\ref{Ihat}) mutually cancel,
and the finite result in this case is given explicitly in section \ref{Scases}.

The original BDS result was derived using the technique of
Mellin-Barnes contour integral representations \cite{BD91}. The
result (\ref{Geomf})-(\ref{Ihat}) or its immediate generalizations,
expressed in terms of generalized hypergeometric functions of two
variables, have been reproduced several times by different authors
using different means \cite{DavDel98,FJT03,SSS03a,SSS03b}. All of
them either contained symmetrizations like (\ref{Sympr}) or
comprised some hidden symmetries in apparently non-symmetric
expressions.

As discussed in the Introduction, for our purposes we needed a
completely different kind of result for the function
$I(m_1^2,m_2^2)$. Thus, we had to calculate the integral
(\ref{I1q}) by using another procedure, not related to that of BDS
or the Feynman parametrization. The way of doing it is described
in the next section.

\section{The alternative calculation}\label{Alter}

Let us return to the integral (\ref{I1q}). Denoting the
denominators of its integrand by $A_-$ and $A_+$, we write it as
$$
I(m_1^2,m_2^2)=\int\frac{d^Dp}{(2\pi)^D}\,\frac{1}{A_-\,A_+}
\quad\quad\mbox{with}\quad\quad
\begin{array}{lp{4,8cm}}
     &$A_-=p^2+m_1^2+1/4-(\bm p\cdot\bm 1)$,\\[-1.7mm]
     &$A_+=p^2+m_2^2+1/4+(\bm p\cdot\bm 1)$.
\end{array}
$$
Using the partial fraction expansion we decompose our integral
into two terms via
$$
I(m_1^2,m_2^2)=\int\frac{d^Dp}{(2\pi)^D}\,\frac{1}{A_-+A_+}
\left(\frac{1}{A_-}+\frac{1}{A_+}\right)\,.
$$
Now, it is useful
to introduce the arithmetic mean $a=(A_-+A_+)/2$ and to express the
denominators $A_-$ and $A_+$ in terms of their mean value $a$ and
deviation from it $b$, as $A_\pm=a\pm b$. Hence,
$$
I(m_1^2,m_2^2)=\int\frac{d^Dp}{(2\pi)^D}\,\frac{1}{2a}
\left(\frac{1}{a-b}+\frac{1}{a+b}\right)\,.
$$
This representation brings into
consideration the mass center $m^2=(m_1^2+m_2^2)/2$ and the
deviation $m_b=(m_2^2-m_1^2)/2\gtrless 0$ of masses $m_1^2$ and
$m_2^2$ from $m^2$. In terms of these values, we have
\bea\nonumber
&&a=p^2+m_a^2\quad\quad\mbox{with}\quad\quad m_a^2=m^2+1/4\,
\quad\quad\mbox{and}\\\nonumber
&&b=(\bm p\cdot\bm 1)+m_b\,.
\eea

The above simple algebraic manipulations provide us with a basis
for further calculations in terms of certain combinations of
$m_1^2$ and $m_2^2$, more appropriate for the parametrization
of final results than the original masses itself.
These allow us to write
$$
I(m_1^2,m_2^2)=\frac{1}{2}\,(I_-+I_+)
$$
with
$$
I_\pm=\int\frac{d^Dp}{(2\pi)^D}\,\frac{1}{p^2+m_a^2}\;\;
\frac{1}{p^2+m_a^2\pm m_b\pm(\bm p\cdot\bm 1)}\,.
$$
Here $m_a^2$
is a manifestly symmetric value related to the mass center $m^2$. On
the other hand, the value $m_b$ is given by the difference of
$m_2^2$ and $m_1^2$. It is not symmetric with respect to the
interchange of masses and thus still not "good". Besides of
dependency on $m^2$, the result must depend on $m_b^2$, not $m_b$.
Our aim is now to reach a manifest realization of the symmetry
$m_1^2\leftrightarrow m_2^2$ at an early stage of the calculation,
not on the level of final results.

As a next step, we perform the angular integration in $I_\pm$ via
$$
I_\pm=K_D\int_0^\infty \frac{p^{D-1}dp}{p^2+m_a^2}\, \int_0^\pi
\frac{d\theta}{\Omega} \frac{\sin^{D-2}\theta}{p^2+m_a^2\pm m_b
\pm p\cos\theta}\,.
$$
Here $K_D$ is a usual geometric factor given by
$K_D=2^{1-D}\pi^{-D/2}/\Gamma(D/2)$.
The normalization factor of the angular integration
$\Omega=\sqrt\pi\,\Gamma(D/2-1/2)/\Gamma(D/2)$ is the value of the
$\theta$ integral without the denominator.
The required result is
$$
I_\pm=K_D\int_0^\infty dp\;
\frac{p^{D-1}}{p^2+m_a^2}\,\; \frac{1}{p^2+m_a^2\pm m_b}\;\;
_2F_1\Big(1,\frac{1}{2};\frac{D}{2};\frac{p^2}{(p^2+m_a^2\pm m_b)^2}\Big).
$$

Hence we obtain, by definition of the Gauss hypergeometric
function, a series representation
\begin{eqnarray}\label{Lastp}
I(m_1^2,m_2^2)&=&\frac{K_D}{2}\sum_{n\ge 0}\frac{ (1/2)_n }{
(D/2)_n } \int_0^\infty dp\, \frac{p^{D-1+2n}}{(p^2+m_a^2)^{2n+2}}
\nonumber\\&\times& \left[ \Big( 1-\frac{m_b}{p^2+m_a^2}
\Big)^{-2n-1}+ \Big( 1+\frac{m_b}{p^2+m_a^2} \Big)^{-2n-1} \right]
\end{eqnarray}
where $(c)_n\equiv\Gamma(c+n)/\Gamma(c)$ is the Pochhammer symbol.
Here, inside of the square brackets we get a
simple realization of the symmetry $m_1^2\leftrightarrow m_2^2$ in
the form of (\ref{Sympr}). This symmetry becomes manifest in the
representation of the content of square brackets in terms of a
Gauss hypergeometric function
(Ref. \onlinecite{SriMan}, p.\ 34, Ref. \onlinecite{PBM3}, p.\ 461, Eq.\ 106)
\be\label{Srm}
\Big( 1-\frac{m_b}{p^2+m_a^2}
\Big)^{-2n-1}+\Big( 1+\frac{m_b}{p^2+m_a^2} \Big)^{-2n-1}=
2\,_2F_1\Big( n+\frac{1}{2},n+1;\frac{1}{2};
\frac{m_b^2}{(p^2+m_a^2)^2} \Big)
\ee
the argument of which depends explicitly on $m_b^2=(m_2^2-m_1^2)^2/4$.

Using equation (\ref{Srm}) we can simply perform the remaining
integration over the radial distance $p$ in (\ref{Lastp}).
Its output is again (see (\ref{Geomf})) given as
$I(m_1^2,m_2^2)=(4\pi)^{-D/2}\Gamma\big(2-D/2\big) \hat
I(m_1^2,m_2^2)$, while for the function $\hat I(m_1^2,m_2^2)$ we
obtain a double series expansion
\begin{equation}\label{Idob}
\hat I(m_1^2,m_2^2)=m_a^{D-4}\sum_{k,n\ge0}\frac{
(\frac{1}{2})_{k+n}(2-D/2)_{2k+n} } {  (\frac{1}{2})_k
(\frac{3}{2})_{k+n}k!n!  }
\Big(\frac{m_b^2}{4m_a^4}\Big)^k\Big(\frac{1}{4m_a^2}\Big)^n\,.
\end{equation}
The same result could be reached directly from equation
(\ref{Lastp}) by using binomial expansions for each of two terms
in the square brackets.

The double series in (\ref{Idob}) does not match directly any item
of the Horn list (see Refs. \onlinecite{EMOT1,SriMan}) of essentially distinct
complete hypergeometric functions of two variables. This leaves a
possibility that it can be reduced to some standard function through
certain algebraic rearrangements.

We see that one of the Pochhammer symbols in the nominator of
(\ref{Idob}) has a doubling of the summation index $k$. This is
rather unusual in the theory of generalized hypergeometric
functions. In order to reduce our double series to a more common
form, one has to get rid of this duplication. This can be achieved
by applying the exact resummation formula (Ref. \onlinecite{SriMan}, p. 100)
$$
\sum_{k,n\ge0}C(n,k)=\sum_{n\ge0}\sum_{k=0}^n C(n-k,k)
$$
along with the identity (Ref. \onlinecite{SriMan}, p. 23)
$$
\frac{1}{(n-k)!}=\frac{(-1)^k(-n)_k}{n!}\,,\quad\quad 0\le k\le n\,.
$$
These manipulations lead us to
\be\label{Isin}
\hat I(m_1^2,m_2^2)=m_a^{D-4}
\sum_{n\ge0}\frac{  (\frac{1}{2})_n (\frac{\ve}{2})_n }{  (\frac{3}{2})_n
n! } \Big(\frac{1}{4m_a^2}\Big)^n \sum_{k=0}^n \frac{  (-n)_k
(\ve/2+n)_k}{(\frac{1}{2})_k k!}\Big(-\frac{m_b^2}{m_a^2}\Big)^k.
\ee
Here we identify the
inner finite sum with $ _2F_1(-n,{\ve/2+n};1/2;-m_b^2/m_a^2)$ and
convert it to a non-terminating hypergeometric function with the
help of the linear transformation formula (see e.g. Ref. \onlinecite{AS}, p.
559, Eqs.\ 15.3.4-5)
\begin{equation}\label{Ltran}
_2F_1(a,b;c;z)=(1-z)^{-a}\,_2F_1\Big(a,c-b;c;\frac{z}{z-1}\Big)\,.
\end{equation}
Thus we get
\be\label{Isn}
\hat
I(m_1^2,m_2^2)=\Big(\frac{\Delta}{4}\Big)^{-\frac{\ve}{2}} \sum_{n\ge0}
\frac{ (\frac{1}{2})_n (\frac{\ve}{2})_n }{  (\frac{3}{2})_n  n! }
\Big(\frac{1}{\Delta}\Big)^n\,
_2F_1\Big(\frac{\ve}{2}+n,\frac{1}{2}+n;\frac{1}{2};
\frac{4m_b^2}{\Delta} \Big)
\ee
where we again encounter the
familiar discriminant $\Delta$ from equation (\ref{Mathcal}). This
time it appears in a compact form as
\begin{equation}\label{Dmamb}
\Delta=4(m_a^2+m_b^2)\,.
\end{equation}

The sum in equation (\ref{Isn}) represents the Appell function
$F_4$  (Ref. \onlinecite{SriMan}, p.\ 53, Ref. \onlinecite{PBM3}, p.\ 413, Eq.\ 5)
whose original definition is \cite{Appell26,Bailey,Slater,SriMan}
\begin{equation}\label{Def4}
F_4(a,b;c,c';x,y)=\sum_{k,n\ge0}
\frac{(a)_{k+n}(b)_{k+n}}{(c)_k(c')_n}\frac{x^k}{k!}\frac{y^n}{n!},\quad
\quad\quad \sqrt{|x|}+\sqrt{|y|}<1.
\end{equation}
Thus, the basic integral
$I(m_1^2,m_2^2)=(4\pi)^{-D/2}\Gamma(\ve/2)\,\hat I(m_1^2,m_2^2)$
is given by
\begin{equation}\label{Iasf4}
\hat I(m_1^2,m_2^2)=\Big(\frac{\Delta}{4}\Big)^{-\frac{\ve}{2}} F_4\Big(
\frac{\ve}{2},\frac{1}{2};\frac{3}{2},\frac{1}{2};
\frac{1}{\Delta},\frac{4m_b^2}{\Delta}\Big)\,.
\end{equation}
For reader's convenience, we recall the notation
$m_b=(m_2^2-m_1^2)/2$ and note an alternative explicit formula
(cf. (\ref{Mathcal})) for the value $\Delta$, which follows
directly from (\ref{Dmamb}):
\be\label{Den}
\Delta=1+2(m_1^2+m_2^2)+(m_2^2-m_1^2)^2\,.
\ee

Let us check the convergence region of the Appell function $F_4$ in (\ref{Iasf4}).
Following (\ref{Def4}), we have to make sure that the combination
\be\label{Conch}
\frac{1}{\sqrt\Delta}+\frac{2|m_b|}{\sqrt\Delta}=
\frac{1+|m_2^2-m_1^2|}{\sqrt\Delta}<1.
\ee
To this end we rewrite the value $\Delta$ as
\bea
\Delta&=&(1+|m_2^2-m_1^2|)^2+2(m_1^2+m_2^2-|m_2^2-m_1^2|)
\nonumber\\\nonumber
&=&(1+|m_2^2-m_1^2|)^2+2\min\{m_1^2,m_2^2\}.
\eea
The last equality gives evidence that $\sqrt\Delta>1+|m_2^2-m_1^2|$, and thus,
the inequality (\ref{Conch}) holds for any $m_1^2,m_2^2>0$.

The equation
(\ref{Iasf4}) is one of our central results. The integral defined
by (\ref{I1q}) is expressed here in terms of a single Appell
function $F_4$. Its variables are given in terms of manifestly
symmetric in $m_1^2$ and $m_2^2$ combinations $\Delta$ and
$m_b^2$. The result for $\hat I(m_1^2,m_2^2)$ is finite and absolutely convergent
in the range $0< D\le 4$ for arbitrary non-vanishing positive masses
$m_1^2$ and $m_2^2$. The limit $D\to 0$ will be considered in
section \ref{Scases} along with further special cases.

\section {A further transformation of the Appell function}

There is a repetition of two parameters in the Appell function
$F_4$ appearing in the equation (\ref{Iasf4}).
This means that the result (\ref{Iasf4}) can be
alternatively expressed in terms of the Horn function $H_3$
(Ref. \onlinecite{SriMan}, p. 57). Moreover, it is reducible to the Appell
function $F_1$ \cite{Appell26,Bailey,Slater,SriMan},
\begin{equation}\label{F1}
F_1(a;b,b';c;x,y)=\sum_{k,n\ge0} \frac{  (a)_{k+n}(b)_k (b')_n
}{(c)_{k+n}}\frac{x^k}{k!}\frac{y^n}{n!}\,,\quad \max\{|x|,|y|\}<1,
\end{equation}
by means of Bailey's \cite{Bailey} reduction formula \cite{SMerr}
\begin{eqnarray}\label{Bail}
&&F_4\Big(
a,b;c,b;\frac{-X}{(1-X)(1-Y)},\frac{-Y}{(1-X)(1-Y)}\Big)
\nonumber\\&& = (1-X)^a(1-Y)^a F_1(a;c-b,1+a-c;c;X,XY).
\end{eqnarray}

Denoting for a while the arguments $1/\Delta$ and $4m_b^2/\Delta$
of the Appell function $F_4$ in (\ref{Iasf4}) as $A^2$ and $B^2$, we find
(cf. Refs. \onlinecite{GN86,GNerr})
\be\label{Xxx}
X=\frac{A^2+B^2-1\pm\sqrt\delta}{2B^2}\,,\quad\quad
Y=\frac{A^2+B^2-1\pm\sqrt\delta}{2A^2} \ee with the new
discriminant \be\label{Mathc}
\delta=(A^2+B^2-1)^2-4A^2B^2%=(B^2-A^2+1)^2-4B^2
=\big[(A+B)^2-1\big]\big[(B-A)^2-1\big],
\ee
astonishingly similar to the function $\Delta$ from equation (\ref{Mathcal}).

The last three equations yield totally unexpected simple expressions.
In particular, we get
$$
\sqrt\delta=4\,\frac{\sqrt{m_1^2m_2^2}}{\Delta}\,.
$$
Choosing the upper $+$ sign in the solutions $X$ and $Y$ from (\ref{Xxx})
and assuming that $m_1>0$ and $m_2>0$ are the principal square roots of
$m_1^2$ and $m_2^2$, we obtain
\begin{equation}\label{Xandy}
X=-\frac{1}{(m_1+m_2)^2}\quad\mbox{and}\quad Y=-(m_1-m_2)^2\,.
\end{equation}
An alternative choice of minus signs in (\ref{Xxx}),
as in Ref. \onlinecite{GN86}, gives $X=-1/(m_1-m_2)^2$ and $Y=-(m_1+m_2)^2$.
This would lead to a much less convenient parametrization of the
function $\hat I(m_1^2,m_2^2)$
owing to the singular behavior of the variable $X$ by equal
parameters $m_1$ and $m_2$.

Thus, the application of the reduction formula (\ref{Bail}) to
the function $F_4$ from (\ref{Iasf4}) leads to the Appell
function $F_1$ with surprisingly simple symmetric arguments. For
$\hat I(m_1^2,m_2^2)$ this yields
\be\label{Iasf1}
\hat I(m_1^2,m_2^2)=\Big(\frac{m_1+m_2}{2}\Big)^{-\ve}
F_1\Bigg(\frac{\ve}{2};1,-\frac{1}{2}+\frac{\ve}{2};\frac{3}{2};
\frac{-1}{(m_1+m_2)^2},\frac{(m_1-m_2)^2}{(m_1+m_2)^2}\Bigg).
\ee
Reintroducing by scaling (\ref{Scaling}) the momentum dependence of the
original Feynman integral (\ref{Idef}) and taking into account the overall
numeric factor (\ref{Geomf}) we obtain a very appealing in its structure result
\bea\label{Iff1}&&
I(p_x;m_1^2,m_2^2)=(4\pi)^{-2+\ve/2}\Gamma\Big(\frac{\ve}{2}\Big)
\\\nonumber&&\quad\quad
\times\Big(\frac{m_1+m_2}{2}\Big)^{-\ve}
F_1\left(\frac{\ve}{2};1,-\frac{1}{2}+\frac{\ve}{2};\frac{3}{2};
\frac{-p_x^2}{(m_1+m_2)^2},\frac{(m_1-m_2)^2}{(m_1+m_2)^2}\right)\,.
\eea

A note is in order here. The last two equations represent the double
series expansions (\ref{F1}) whose domain of absolute convergence is shrunk with
respect to the original integral's validity range,
$m_1>0$, $m_2>0$, and arbitrary $p_x$.
Indeed, the definition (\ref{F1})
of the Appell function $F_1$ requires $m_1+m_2>1$ for the first variable of the
scaling function $\hat I(m_1^2,m_2^2)$ in (\ref{Iasf1}), or
$p_x/(m_1+m_2)<1$ for the whole integral (\ref{Iff1}).
Nevertheless, this is not an essential drawback.
To get rid of it, one can apply the linear transformation
(Ref. \onlinecite{Appell26} p. 30, Eq. (5$_3$),
Ref. \onlinecite{Slater} p. 218, Eq. (8.3.3))
\be\label{Ltf1}
F_1(a;b,b';c;u,v)=(1-u)^{-a}F_1\Big(a;c-b-b',b';c;\frac{u}{u-1},
\frac{u-v}{u-1}\Big)
\ee
to the Appell function $F_1$ in (\ref{Iasf1}).
This provides the necessary analytic continuation for this function,
very similar to
that given by the linear transformation (\ref{Ltran}) for the Gauss hypergeometric
function: The first argument of $F_1$ behaves in exactly the same way as the
variable $z$ in (\ref{Ltran}). Hence we obtain, for any $m_1$, $m_2>0$,
\bea\label{Jasf1}
\hat I(m_1^2,m_2^2)&=&2^\ve\left[1+(m_1+m_2)^2\right]^{-\frac{\ve}{2}}\\\nonumber
&\times& F_1\left(\frac{\ve}{2};1-\frac{\ve}{2},\frac{\ve}{2}-\frac{1}{2};\frac{3}{2};
\frac{1}{1+(m_1+m_2)^2},\frac{1+(m_1-m_2)^2}{1+(m_1+m_2)^2}\right)\,.
\eea
The convergence range of this double series expansion now coincides with the
initial validity domain of the original integral (\ref{Idef}).

Although with a reduced domain of convergence,
the compact expressions (\ref{Iasf1}) or (\ref{Iff1}) can be used in
further calculations, where the needed analytic continuation is done at
some later stage or is provided by the extended convergence region of
final results. Examples of this kind will be given in the following section.
Also, the Appell function $F_1$ is presently a built-in function in the
Mathematica \cite{math},
and its analytic continuation is automatically provided by this program.

\section{Special cases}\label{Scases} %, and (\ref{Jasf1})

Our general results (\ref{Iasf4}) and (\ref{Iasf1})-(\ref{Iff1}) can be checked
by specializing to several integer space dimensions $D$.
Here we encounter mathematical simplifications allowing us to get
the expressions for $\hat I(m_1^2,m_2^2)$ in terms of elementary
functions. Some of them have already been calculated by other means
before.

At $D=1$, the integrals (\ref{Idef}) or (\ref{I1q}) are elementary, and
a short calculation using Mathematica \cite{math} yields
\begin{equation}\label{Id0}
\hat I(m_1^2\left.,m_2^2)\right|_{D=1}=2\,\frac{m_1+m_2}{m_1 m_2}\,
\frac{1}{1+(m_1+m_2)^2}.
\end{equation}
The same result immediately follows from (\ref{Iasf1}): With $\ve=3$,
the Appell function $F_1$ simply decouples there into two geometric progressions.

At $D=2$, the integral $\hat I(m_1^2,m_2^2)$ has been calculated
in Ref. \onlinecite{SPD05}, in the context of the $1/N$ expansion for
uniaxial ($m=1$) Lifshitz points in three-dimensional ($d=3$) space. In this
special case we have $\ve=2$, and we obtain from the equation (\ref{Iasf4})
\begin{equation}\label{Iasd3}
\hat I(m_1^2\left.,m_2^2)\right|_{D=2}=\frac{4}{\Delta}\, F_4\Big(
1,\frac{1}{2};\frac{3}{2},\frac{1}{2};
\frac{1}{\Delta},\frac{4m_b^2}{\Delta} \Big)\,.
\end{equation}
This simplified Appell function reduces through the relation
(Ref. \onlinecite{Bailey}, p.\ 102)
\begin{eqnarray}\label{Bail4}
&&F_4\Big( a,b;1+a-b,b;\frac{-X}{(1-X)(1-Y)},\frac{-Y}{(1-X)(1-Y)}
\Big) \nonumber\\&&= (1-Y)^a\,
_2F_1\Big(a,b;1+a-b;-X\frac{1-Y}{1-X}\Big)
\end{eqnarray}
to a Gaussian hypergeometric function with parameters 1,1/2;3/2
and positive argument $w=(1+(m_1-m_2)^2)/(1+(m_1+m_2)^2)<1$.
It is known (e.g.\ Ref. \onlinecite{AS}, p.\ 556)
to represent a logarithmic function. Thus we obtain
\begin{equation}\label{D3ln}
\hat I(m_1^2\left.,m_2^2)\right|_{D=2}=\frac{2}{\sqrt\Delta}
\ln\frac{1+\sqrt w} {1-\sqrt w}
=\frac{2}{\sqrt\Delta}\ln\frac{1+m_1^2+m_2^2+\sqrt{\Delta}}{2m_1m_2}\,.
\end{equation}
This special result comes as well from (\ref{Iasf1}) through the reduction
formula (Ref. \onlinecite{Bailey}, p.\ 79)
\begin{equation}\label{Bail1}
F_1( a;b,b';b+b';u,v ) = (1-v)^{-a}\,_2F_1\Big(a,b;b+b';\frac{u-v}{1-v}\Big)
\end{equation}
followed by the linear transformation (\ref{Ltran})
for the resulting Gauss hypergeometric function. The latter transformation
converts the argument $z\equiv(u-v)/(1-v)=-(1+(m_1-m_2)^2)/(4m_1m_2)$,
which blows up for small values of
$m_1$ and $m_2$, again to the safe combination $w<1$.
This is a counterpart of the analytical continuation carried out
before through the transformation (\ref{Ltf1}) for the Appell function $F_1$.

The last expression in (\ref{D3ln}) agrees with the integrand of
$I(1,q)$ from equation (70) of Ref. \onlinecite{SPD05} after the shift of the
integration variable via $q'\to q'-q/2$ and identifications
$m_1=(q'-q/2)^2$ and $m_2=(q'+q/2)^2$. After some work, it can also be seen to
be equivalent with the formula (4.3) of Ref. \onlinecite{DavDel98}.

At $D=3$, the dimensional parameter $\ve=1$, and in the
Appell function $F_1$ from (\ref{Iasf1}) one of the nominator
parameters vanishes. Thus, the function $F_1$ reduces to an ordinary series expansion,
and we get
\begin{equation}\label{D3a}
\hat I(m_1^2\left.,m_2^2)\right|_{D=3}=2 \arctan(m_1+m_2)^{-1}
\end{equation}
in agreement with Ref. \onlinecite{Raj96}, Eq. (A.2) \cite{Rajnote}
and Ref. \onlinecite{DavDel98}, Eq. (4.6).

When $D=4$, the value of $\ve$ vanishes, and both the equations
(\ref{Iasf4}) and (\ref{Iasf1}) yield the trivial result $\hat
I(m_1^2\left.,m_2^2)\right|_{D=4}=1$. Now, of interest is the
first-order term of a small-$\ve$ expansion of $\hat
I(m_1^2,m_2^2)$ taken at $D=4-\ve$. Its calculation will yield the
essential non-trivial contribution to the finite part of the whole
integral $I(m_1^2,m_2^2)$ given by equations
(\ref{I1q})-(\ref{Geomf}). This term has been calculated
previously by BDS \cite{BDS96} using the Feynman parametrization.

Expanding the result (\ref{Iasf4}) in small $\ve$ we obtain (see Appendix A)
\bea\nonumber
\hat I(m_1^2,m_2^2)&=&1+\ve+
\frac{\ve}{4}\sqrt\Delta\,\ln\frac{1+m_1^2+m_2^2-\sqrt\Delta}
{1+m_1^2+m_2^2+\sqrt\Delta}
+\frac{\ve}{4}(m_2^2-m_1^2)\ln\frac{m_2^2}{m_1^2}
\\\label{Epsi}&-&
\frac{\ve}{4}\ln(m_1^2m_2^2)+O(\ve^2)
\eea
in agreement with Refs. \onlinecite{BDS96} and \onlinecite{DavDel98}.

The limiting case $D=0$ is somewhat special. It can be well illustrated
by turning to the "double" integral (\ref{LPint}) stemming from the
Lifshitz-point theory.
Let us accept again that $D=d-m$ while the $d$-dimensional space
is split into mutually complementing $m$- and $D$-dimensional subspaces.
Then, by inspecting (\ref{LPint}), it becomes evident that when the
dimension $D$ shrinks to zero, we must remain with an $m$ dimensional
integral over $\bm q$ where no trace of $D$ dimensional integration remains.
This implies that in the limit $D\to 0$,
the result of the integration over $\bm p$
must yield its integrand at $\bm p=\bm p_x=0$.
Physically relevant limiting regimes of this kind have been considered in
Refs. \onlinecite{SD01,SPD05}.
For the integral $I(p_x;m_1^2,m_2^2)$ from (\ref{Idef})
this means that
\be\label{D0}
I(p_x;m_1^2\left.,m_2^2)\right|_{D=0}=\frac{1}{m_1^2m_2^2}\,.
\ee
This is easily reproduced from equation (\ref{Iff1}).
At $p_x=0$, the first argument of $F_1$ vanishes, and it reduces to a
Gauss hypergeometric function yielding
$$
I(p_x=0;m_1^2,m_2^2)=(4\pi)^{-2+\ve/2}\Gamma\Big(\frac{\ve}{2}\Big)
\frac{2}{2-\ve}\frac{m_1^{2-\ve}-m_2^{2-\ve}}{m_1^2-m_2^2}
$$
in agreement, up to the normalization of the integral, with Ref. \onlinecite{BD91}.
By setting $\ve=4$ here, we obtain (\ref{D0}).
Note that this correct limit could not be obtained by setting $D=0$
directly in the scaling
function (\ref{Iasf1}).There, no vanishing of the first argument
in $F_1$ occurs, which was achieved by restoring the external momentum dependence
in (\ref{Iff1}).

\section{Reduction relations for the Appell functions}

By equating the results (\ref{Iasf4}), (\ref{Iasf1}), or (\ref{Jasf1}) for the
integral $\hat I(m_1^2,m_2^2)$ with its BDS expression
(\ref{Ihat}) we obtain apparently new reduction relations for
the involved Appell functions. Thus, for the function $F_4$ we get
\bea\label{Tot1}
&&F_4\Big(\alpha,\frac{1}{2};\frac{3}{2},\frac{1}{2}; \frac{1}{
\Delta},\frac{(m_2^2-m_1^2)^2}{\Delta} \Big)
=4^{-\alpha}\,\frac{\Gamma^2(1-\alpha)}{\Gamma(2-2\alpha)}\sqrt\Delta
\\&&\nonumber
+\left(\frac{\Delta}{4m_1^2}\right)^\alpha
\frac{1+m_2^2-m_1^2-\sqrt\Delta}{2(1-\alpha)}\,
_2F_1\Big( 1,\alpha;2-\alpha; -\,\frac{
(1+m_2^2-m_1^2-\sqrt\Delta)^2 }{ 4m_1^2 } \Big)
\\&&\nonumber
+\left(\frac{\Delta}{4m_2^2}\right)^\alpha
\frac{1+m_1^2-m_2^2-\sqrt\Delta}{2(1-\alpha)}\,
_2F_1\Big( 1,\alpha;2-\alpha; -\,\frac{
(1+m_1^2-m_2^2-\sqrt\Delta)^2 }{ 4m_2^2 } \Big)
\eea
where a parameter $\alpha$ stands in place of  $2-D/2=\ve/2$.
The relation (\ref{Tot1}) was derived for arbitrary real positive masses
$m_1^2$ and $m_2^2$.
This is in conformity with the standard definition of the convergence region of
the Appell function $F_4$, $(1+|m_2^2-m_1^2|)/\sqrt\Delta <1$,
as discussed at the end of the section \ref{Alter}.
Once again we recall that $\Delta=1+2(m_1^2+m_2^2)+(m_2^2-m_1^2)^2$.

The parameter $\alpha$ was originally constrained to the interval
$0\le\alpha<2$ by
physical applicability range of the integral (\ref{I1q}).
In (\ref{Tot1}), this limitation can be removed.
The relation (\ref{Tot1}) holds for negative values of $\alpha$,
where the original integrals are divergent in the ultraviolet,
or for $\alpha> 2$ where the integrals would be negative dimensional.
We have numerical evidence that (\ref{Tot1}) is valid for integer numbers
$\alpha\ge2$ where the hypergeometric functions become singular and one
might beware of possible exceptions.
This is in agreement with the smooth dependence of $F_4$ on $\alpha$
that indicates no peculiarities for positive integers.
Moreover, in the relation (\ref{Tot1}),
the parameter $\alpha$ can be considered to extend to
complex values, along with the variables $m_1^2$ and $m_2^2$.

Eliminating the parameters $m_1^2$ and $m_2^2$ in favor of
variables that appear in the function $F_4$ we express the last relation
in a standard fashion,
\bea\label{Totxy}
&&F_4\Big(\alpha,\frac{1}{2};\frac{3}{2},\frac{1}{2}; x,y\Big)
=4^{-\alpha}\,\frac{\Gamma^2(1-\alpha)}{\Gamma(2-2\alpha)}\;\frac{1}{\sqrt x}
\\&&\nonumber
+\left[1-(\sqrt x+\sqrt y)^2\right]^{-\alpha}
\frac{\sqrt x+\sqrt y-1}{2(1-\alpha)\sqrt x}\,
_2F_1\Big( 1,\alpha;2-\alpha;\frac{\sqrt x+\sqrt y-1}
{ \sqrt x+\sqrt y+1 } \Big)
\\&&\nonumber
+\left[1-(\sqrt x-\sqrt y)^2\right]^{-\alpha}
\frac{\sqrt x-\sqrt y-1}{2(1-\alpha)\sqrt x}\,
_2F_1\Big( 1,\alpha;2-\alpha;\frac{\sqrt x-\sqrt y-1}
{ \sqrt x-\sqrt y+1 } \Big)\,.
\eea

Comparing the equations (\ref{Iasf1}) and (\ref{Ihat})
we write down a similar reduction relation for the Appell function $F_1$,
\bea\label{Tot2}
&&F_1\Big(\alpha;1,\alpha-\frac{1}{2};\frac{3}{2};-p^2,q^2\Big)
=4^{-\alpha}\,\frac{\Gamma^2(1-\alpha)}{\Gamma(2-2\alpha)}\,(p^2)^{-1+\alpha}\,
s^{\frac{1}{2}-\alpha}
\\&&\nonumber
+\frac{(1+q)^{-2\alpha}}{2(1-\alpha)p^2}\left(p^2-q-\sqrt s\right)\,
_2F_1\Big( 1,\alpha;2-\alpha; -\,\frac{
(p^2-q-\sqrt s)^2 }{ p^2(1+q)^2 } \Big)
\\&&\nonumber
+\frac{(1-q)^{-2\alpha}}{2(1-\alpha)p^2}\left(p^2+q-\sqrt s\right)\,
_2F_1\Big( 1,\alpha;2-\alpha; -\,\frac{
(p^2+q-\sqrt s)^2 }{ p^2(1-q)^2 } \Big)\,,
\eea
where
$$
s=(1+p^2)(p^2+q^2).
$$
By writing in (\ref{Tot2}) the variables of $F_1$ as $-p^2$ and $q^2$
we stress that this function
is even in $p$ and $q$. So is the combination on the right, too.
But here, each of terms with a Gauss hypergeometric function
contains an explicit dependence on $q$ and is not even in this variable.
All the odd powers of $q$ must cancel in the whole combination through the
symmetrization $f(q)+f(-q)$ present here. Its role is analogous
to that of (\ref{Sympr}), implemented in (\ref{Ihat}) and (\ref{Tot1}).

The Appell function $F_1$ from the relation (\ref{Tot2})
exists for any real or complex values of the
parameter $\alpha$. It is absolutely convergent for
$p^2$ and $q^2<1$. These variables can also be both real and complex.
The convergence region of the function $F_1$ can be extended by using
the transformation formulas like that in (\ref{Ltf1}). A further
reduction relation for the Appell function $F_1$ can be written in
a similar way by using the formula (\ref{Jasf1}).

\section{Concluding remarks}

In this paper we have discussed the functional form of a standard
but non-trivial one-loop Feynman integral (\ref{Idef}) with non-vanishing masses
($m_1^2$ and $m_2^2$) and external momentum in $D$ spatial dimensions.
Previously the results for this Feynman integral have been obtained
by Berends, Davydychev, and Smirnov
\cite{BDS96} and reproduced later \cite{DavDel98,FJT03,SSS03a,SSS03b}
in several equivalent forms.
The BDS result contained a linear combination of two
Gauss hypergeometric functions neither of which was symmetric with respect to
the interchange of the masses $m_1^2$ and $m_2^2$.
The correct symmetric part of this combination had to remain
after very complicated and obscured cancelations of non-symmetric terms.

Here we represent the new explicit results for the Feynman integral (\ref{Idef})
in terms of Appell functions, manifestly symmetric with respect to the masses
$m_1^2$ and $m_2^2$.
The advantage of our results over other representations
is that they are expressed by a \emph{single}
function with completely symmetric and rather simple arguments.
In equation (\ref{Iasf4}) the result is given in terms of $F_4$,
"perhaps the most intractable Appell function" \cite{Exton95}.
A further transformation of this function led us to
elegant results ({\ref{Iasf1}})-({\ref{Iff1}}) and ({\ref{Jasf1}})
expressed in terms of the Appell function $F_1$ with
extremely simple and symmetric arguments.

The knowledge of new functional forms of the integral (\ref{Idef})
gave us the possibility to derive the equalities relating the involved Appell
functions to previously known combinations of Gauss hypergeometric functions.
The way of producing the new mathematical identities by evaluating
some integrals in different ways is not new. It was successfully employed
by Inayat-Hussain \cite{IH87,IH87a} (see also Ref. \onlinecite{BSr90})
while considering certain Feynman integrals
arising in phase transition theory, or by Srivastsva, Glasser, and Adamchik
\cite{SrGA00} while studying the Riemann Zeta function.

Apart from equations (\ref{Tot1})-(\ref{Tot2}) for
$F_4$ and $F_1$, similar reduction relations can be derived also for
the Appell functions $F_3$ and $F_2$ since $F_1$ can always be
expressed in terms of one of these latter functions \cite{Appell26,Bailey}.
Of special interest would be the relations involving $F_2$ in view of attention
attracted by this function in the recent mathematical literature
\cite{Tar98,Tar03,Oppsss05}.

\section*{Acknowledgment}
It is my  pleasure to thank H. W. Diehl for his hospitality at
the Duisburg-Essen university where the present work has been initiated.
The partial support by the Deutsche Forschungsgemeinschaft (DFG)
via  the Leibniz program Di 378/3 is gratefully acknowledged.

\appendix
\section{The epsilon expansion of $\hat I(m_1^2,m_2^2)$}  \label{Ascf}

Let us consider the equation (\ref{Iasf4}) at small $\ve$.
For a while, we denote the arguments
$1/\Delta$ and $4m_b^2/\Delta$ of the Appell function $F_4$ as $x$
and $y$, respectively. We use the definition (\ref{Def4}) of $F_4$
and split there the $k=0$ term. Denoting it by $\sigma_0$ we have
\be\label{Sinul}
\sigma_0=\sum_{n\ge 0}\frac{(\ve/2)_n}{n!}\,y^n= 1+\frac{\ve}{2}
\sum_{n\ge 1}\frac{y^n}{n}+O(\ve^2)=1-\frac{\ve}{2}\ln(1-y)+O(\ve^2)
\ee
since
\be\label{Poch}
(a)_n=a(n-1)!+O(a^2) \quad\mbox{if}\quad n\ge 1.
\ee
After a rearrangement of Pochhammer symbols via $(a)_{n+k}=(a)_k(a+k)_n$,
the rest of the double series expansion can be written as
$$
\sigma_1=\sum_{k\ge 1}\frac{(\ve/2)_k(1/2)_k}
{(3/2)_k}\,\frac{x^k}{k!}\,_2F_1(\ve/2+k,1/2+k;1/2;y).
$$
Again we use the property (\ref{Poch}) of $(\ve/2)_k$ to obtain
$$
\sigma_1=\frac{\ve}{2}\sum_{k\ge 1}\frac{(1/2)_k}
{(3/2)_k}\,\frac{x^k}{k}\,_2F_1(k,1/2+k;1/2;y)+O(\ve^2).
$$
Here, the hypergeometric function reduces to an algebraic function
(Ref. \onlinecite{SriMan}, p.\ 34, Ref. \onlinecite{PBM3}, p.\ 461, Eq.\ 106)
through the same relation that was used in (\ref{Srm}). Thus we get
$$
\sigma_1=\frac{\ve}{4}\sum_{k\ge 1}\frac{(1/2)_k}
{(3/2)_k}\,\frac{x^k}{k}\left[(1+\sqrt y)^{-2k}+(1-\sqrt y)^{-2k}
\right]+O(\ve^2).
$$
This formula implies that, if we introduce the series expansion
\be\label{Sidef}
\sigma(z)\equiv\sum_{k\ge 1}\frac{(1/2)_k}{(3/2)_k}\,\frac{z^k}{k},
\ee
the function $\sigma_1$ is given by
\be\label{Simax}
\sigma_1=\frac{\ve}{4}\left[\sigma (X_+)+\sigma (X_-)\right]+O(\ve^2)
\quad\quad\mbox{where}\quad\quad
X_\pm\equiv\frac{x}{(1\pm\sqrt y)^2}.
\ee
By shifting the summation index in (\ref{Sidef}), we can write $\sigma(z)$
as a a generalized hypergeometric function $_3F_2$:
$$
\sigma(z)=\frac{z}{3}\;_3F_2\Big(\frac{3}{2},1,1;\frac{5}{2},2;z\Big).
$$
This can be expressed via (Ref. \onlinecite{PBM3}, p. 519, Eq. 366) in terms of
logarithmic functions:
\be\label{Siz}
\sigma(z)=2-\frac{1}{\sqrt z}\,\ln\frac{1+\sqrt z}{1-\sqrt z}-\ln(1-z)\,.
\ee
The same result could be reached by noticing that the derivative of $\sigma(z)$
is given by
$$
\sigma'(z)=\sum_{k\ge 1}\frac{(1/2)_k}{(3/2)_k}\,z^{k-1}=
\frac{1}{3}\;_2F_1\Big(\frac{3}{2},1;\frac{5}{2};z\Big)=
\frac{1}{z}\left[\frac{1}{2\sqrt z}\,\ln\frac{1+\sqrt z}{1-\sqrt z}-1\right],
$$
where the last equality follows from Ref. \onlinecite{PBM3}, p. 477, Eq. 157,
and integrating back $\sigma'(z)$ with respect to $z$.

The summary of the above calculation is that the first-order $\ve$ expansion
of the Appell function $F_4$ from (\ref{Iasf4}) reads
$$
F_4\Big(\frac{\ve}{2},\frac{1}{2};\frac{3}{2},\frac{1}{2};x,y\Big)=
{1}-\frac{\ve}{2}\ln(1-y)+
\frac{\ve}{4}\left[\sigma\Big(\frac{x}{(1+\sqrt y)^2}\Big)
+\sigma\Big(\frac{x}{(1-\sqrt y)^2}\Big)\right]+O(\ve^2)
$$
where the function $\sigma(z)$ is given explicitly by (\ref{Siz}).

Finally, the small-$\ve$ expansion of the complete formula (\ref{Iasf4})
appears in the equation (\ref{Epsi}) of the main text.


\begin{thebibliography}{35}
\bibitem[{GNe()}]{GNerr}
\bibinfo{note}{The denominator of $y$ in equation (2.4b) of Ref. \onlinecite{GN86} should
  read $2a^2$.}
\bibitem[{Raj()}]{Rajnote}
\bibinfo{note}{The explicit expressions for analytically computable three-loop
  {F}eynman integrals at ${D}=3$ and $m_1^2=m_2^2=m^2$ used in Ref. \onlinecite{Raj96},
  have been given before in Ref. \onlinecite{BSh92}.}
\end{thebibliography}
\end{document}